\documentclass[draft,jgrga]{agutex}

 \usepackage{lineno}
 \linenumbers*[1]

  \usepackage[dvipdf]{graphicx}

  \setkeys{Gin}{draft=false}
\authorrunninghead{YERMOLAEV ET AL.}

\titlerunninghead{INFLUENCE OF DRIVER TYPE ON THE STORM DURATIONS}

\begin{document}

%% ------------------------------------------------------------------------ %%
%
%  TITLE
%
%% ------------------------------------------------------------------------ %%

\title{%[Estimation of] 
Influence of the interplanetary driver type on the durations of main and recovery phases of magnetic storms 
%Or 
%[Estimation of] probability distribution function of extreme magnetic storms
}
%
% e.g., \title{Terrestrial ring current:
% Origin, formation, and decay $\alpha\beta\Gamma\Delta$}
%

%% ------------------------------------------------------------------------ %%
%
%  AUTHORS AND AFFILIATIONS
%
%% ------------------------------------------------------------------------ %%

%Use \author{\altaffilmark{}} and \altaffiltext{}

% \altaffilmark will produce footnote;
% matching \altaffiltext will appear at bottom of page.

% \authors{R. C. Bales,\altaffilmark{1}
% E. Mosley-Thompson,\altaffilmark{2} R. Williams,\altaffilmark{3}
% J. R. McConnell\altaffilmark{4}, and Francesco Visconti\altaffilmark{5}} 

\authors{Yu. I. Yermolaev, \altaffilmark{1}
I. G. Lodkina, \altaffilmark{1} 
N. S. Nikolaeva , \altaffilmark{1} 
M. Yu. Yermolaev \altaffilmark{1}
}

\altaffiltext{1}{Space Plasma Physics Department, Space Research Institute, 
Russian Academy of Sciences, Profsoyuznaya 84/32, Moscow 117997, Russia. 
(yermol@iki.rssi.ru)}

%\altaffiltext{1}{Department of Hydrology and Water Resources,
%University of Arizona, Tucson, Arizona, USA.}

%\altaffiltext{2}{Department of Geography, Ohio State University,
%Columbus, Ohio, USA.}

%\altaffiltext{3}{Department of Space Sciences, University of
%Michigan, Ann Arbor, Michigan, USA.}

%\altaffiltext{4}{Division of Hydrologic Sciences, Desert Research
%Institute, Reno, Nevada, USA.}

%\altaffiltext{5}{Dipartimento di Idraulica, Trasporti ed
%Infrastrutture Civili, Politecnico di Torino, Turin, Italy.}

%% ------------------------------------------------------------------------ %%
%
%  ABSTRACT
%
%% ------------------------------------------------------------------------ %%

% >> Do NOT include any \begin...\end commands within
% >> the body of the abstract.

\begin{abstract}

We study durations of main and recovery phases of magnetic storms induced 
by different types of large-scale solar-wind streams (Sheath, magnetic cloud (MC), Ejecta and CIR) 
on the basis of OMNI data base during 1976-2000. 
Durations of both main and recovery phases depend on types of interplanetary drivers. 
On the average, duration of main phase of storms induced by compressed regions 
(CIR and Sheath) is shorter than by MC and Ejecta 
while duration of recovery phase of CIR- and Sheath-induced storms is longer. 
Analysis of durations of individual storms shows that durations of main and recovery phases 
anti-correlate for CIR- and Sheath-induced storms and there is not dependence between them 
for (MC+Ejecta)-induced storms.

\end{abstract}

%% ------------------------------------------------------------------------ %%
%
%  BEGIN ARTICLE
%
%% ------------------------------------------------------------------------ %%

% The body of the article must start with a \begin{article} command
%
% \end{article} must follow the references section, before the figures
%  and tables.

\begin{article}

%% ------------------------------------------------------------------------ %%
%
%  TEXT
%
%% ------------------------------------------------------------------------ %%

\section{Introduction}

One of the main problems (aims) of the solar-terrestrial physics is the investigation of connection 
between interplanetary conditions and 
magnetospheric activity including magnetic storms. As has been shown by first works %(see, i.e., papers 
by 
\cite{Dungey1961,FairfieldCahill1966,RostokerFalthammar1967,Russelletal1974,Burtonetal1975,Akasofu1981}
% – Yermolaev_AnnGeo2010) 
%and references therein) 
the most important parameter leading to geomagnetic disturbances and, in particular, 
to magnetic storm generation is negative 
(southward) Bz component of interplanetary magnetic field (IMF) or $Ey = Vx Bz$ component 
of electric field where $Vx$ is radial component of solar wind velocity. 
These papers have shown that %value of 
the magnetic storm value (minimal $Dst$ index) depends also on the duration of southward $Bz$ component 
(or $Ey$) action.  
These results allowed 
\cite{Burtonetal1975} 
to describe the dynamics of magnetic storm  by formula $d Dst/dt = a Ey –- Dst/\tau$, 
e.g. a current value of magnetic storm is result of competing general processes: excitation and relaxation. 
Each process depends on many parameters and has characteristic times and durations, and studies of them can give us 
an additional information about physical mechanisms of these processes. 

A numerous papers studied durations of main phase of magnetic storms. Authors of numerous papers 
(see, i.e., 
\cite{Russelletal1974,GonzalezTsurutani1987,Tsurutaniaetal1992,Gonzalezetal1994,TsurutaniGonzalez1995,Wangetal2003,Nikolaevaetal2011} and references therein) 
obtained empirical durations of corresponding IMF for threshold of southward $Bz$ component which are necessary 
for generation of storm of given value. 

For example,  
\cite{Wangetal2003} 
obtained results that southward field component $Bz \ge 3$ nT with a duration of $\Delta t \ge 1$ h results 
in moderate magnetic storms ($Dst_{min} \le -–50$ nT) and 
threshold values of $Bz \ge 6$ nT with a duration of $\Delta t \ge 2$ h in strong magnetic 
storms ($Dst_{min} \le –-100$ nT), and 
these results differ somewhat from the results of previous papers by %authors 
\cite{Russelletal1974,GonzalezTsurutani1987,Gonzalezetal1994}. 
Papers mentioned above analyzed duration of southward component which is enough to generate magnetic storms  
but duration of main phase was studied less intensively and 
it was shown that duration of main phase may be from 2 h to 1 day 
(see., i.e.,  
\citep{YokoyamaKamide1997,Vieiraetal2004,Vichareetal2005,GonzalezEcher2005,Yermolaevetal2007,Hutchinsonetal2011,Nikolaevaetal2012} 
and references therein). 

Though it is well known that dynamics of magnetic storms depends on type interplanetary drivers 
(see, i.e., papers by  
\cite{BorovskyDenton2006,Yermolaevetal2010,Guoetal2011,LiemohnKatus2012,Crameretal2013}
and references therein),  
majority of previous works did not make selection of types of solar wind streams which generated storms. 
In another papers selection was either performed only for limited type of
solar wind or for complex of types. For instance, in recent paper by 
\cite{Hutchinsonetal2011} 
there is a definition of CME type 
in  caption of Figure 2: 
''A typical coronal mass ejection (CME) trace seen in ACE OMNI data, with simultaneous increases 
in all components: interplanetary magnetic field (IMF) magnitude, solar wind (SW) speed, pressure, 
density, and temperature.'' 
This is the definition of compression region (Sheath) before body of interplanetary CME (ICME). 
It is unclear, whether in the article authors only analyze Sheath before ICME body (magnetic cloud or Ejecta) 
or the complex 
including Sheath and ICME body. 
It is necessary to note that there are significant differences between storms driven by Sheath 
and ICME:   
the storms driven by Sheath regions have stronger magnetotail field stretching, larger asymmetry 
in the inner magnetosphere field configuration, and larger asymmetric ring current, 
while the ring current enhancement is stronger 
in the storms driven by magnetic clouds (see, i.e., 
\citep{Huttunenetal2002, Huttunenetal2006,Pulkkinenetal2007,Guoetal2011},  and references therein). 
In our previous paper 
\citep{Nikolaevaetal2012} 
we showed that on average, the smallest duration of the main phase 
($\Delta T \sim 5.5$ h) is observed for magnetic storms related to Sheath regions $Sh_{E}$ 
and $Sh_{MC} + Sh_{E}$, and the largest duration 
of the main phase ($\Delta T \sim 8.5–-9$ h), for magnetic storms caused by their Ejecta and $MC + Ejecta$ bodies. 
These values of duration are twice less, than values presented in paper by 
\cite{Mustajab2013} 
and they are in good agreement with mean results of paper by 
\cite{Hutchinsonetal2011}. %Hutchinson et al (2011).

Recovery phase of magnetic storms and processes responsible for decay of ring current are 
a subject of numerous  papers (see e.g., 
\cite{Daglisetal1999,Keikaetal2006,XuDu2010,OBrienMcPherron2000,Feldsteinetal2000,MonrealLlop2008,Yermolaevetal2012}.
%Daglis et al., 1999; Keika et al., 2006; Xu and Du, 2010, O'Brien
%and McPherron, 2000; Feldstein et al., 2000; Monreal MacMahon and Llop-Romero, 2008; Xu and Du, 2010, 
%Yermolaev et al., 2012 
and references therein) . 
In our recent paper 
\citep{Yermolaevetal2012} 
we show that the recovery phase Dst variations 
depend on type of interplanetary drivers inducing magnetic storms 
(e.g., magnetosphere remembers type of driver during recovery phase) and 
durations of recovery phase for storms induced by CIR and Sheath are longer 
than for storms induced by MC and Ejecta.

Dependences of durations of main and recovery phases of storms on the storm value ($Dst_{min}$) 
are studied in several papers (
%Yokoyama and Kamide, 1997, Hutchinson et al (2011)
\cite{YokoyamaKamide1997,Hutchinsonetal2011}). 
\cite{YokoyamaKamide1997} 
obtained that there is correlation between duration of both phases and Dstmin. 
Results of paper by 
\cite{Hutchinsonetal2011} 
confirm previous results for recovery phase 
while they have more complicated (nonmonotonic) form for main phase: 
at initial stage duration increases with growth of storm value and then it decreases. 
In accordance of our results  (
\cite{Yermolaevetal2012}), 
for storms generated by MC and Sheath, there are 2 classes of storms: 
duration of recovery phase and $|Dst_{min}|$ correlate for short storm duration and 
they have inverse correlation for long storm duration while for storms generated by CIR and Ejecta, 
durations of recovery phase of storm slightly depend on $Dst_{min}$. 

In this paper we study the durations of main and recovery phases of magnetic storms induced by different
interplanetary drivers, relations between durations of both phases as well as their dependence on magnetic storm value.

\section{Methods} 

This paper is a continuation of our previous paper 
[Yermolaev et al., 2012] and 
we use the same data bases (OMNI data of solar wind and interplanetary magnetic field parameters 
(see http://omniweb.gsfc.nasa.gov) [King and Papitashvili, 2005] and 
data on Dst index (see http://wdc.kugi.kyoto-u.ac.jp/index.html for period of 1876-2000) and 
the same method of classification of large-scale solar wind streams 
(see our catalog of large-scale interplanetary events for period of 1976–-2000 in web site 
ftp://www.iki.rssi.ru/pub/omni 
[Yermolaev et al., 2009]). 
We study separately 798 moderate and strong magnetic storms with $-100 < Dst_{min} \le -50$ nT and 
$Dst_{min} \le -100$ nT, respectively. 
Our analysis for period of 1976–2000 showed that 145 magnetic storms  have been generated by CIR, 
96 storms by Sheath, 62) by MC and 161 by Ejecta. 
The sources of other 334 magnetic storms  are indeterminate and these storms are indeterminate 
and indicated as (IND) 
[Yermolaev et al., 2010a, 2012]. About 20\% of storms were multistep ones 
during main and  recovery phases and these storms were excluded from analysis.

We use the same method to measure duration of recovery phase: 2 times when $Dst$ index achieves 
levels of 1/2 and 1/3 
from minimum $Dst$ index as criteria of time of recovery
phase termination, and analyze two durations $\Delta t_{1/2}$ and 
$\Delta t_{1/3}$, i.e. time intervals from $Dst_{min}$ up to $(1/2)Dst_{min}$ 
($\Delta t_{1/2} = t((1/2)Dst_{min}) - t(Dst_{min})$) and $(1/3)Dst_{min}$ 
($\Delta t_{1/3} = t((1/3)Dst_{min}) - t(Dst_{min})$), respectively. 
Comparison of two data sets corresponding to $\Delta t_{1/2}$ and $\Delta t_{1/3}$ allows us to 
make conclusions about dynamics of $Dst$ index during storm recovery phase.
%Dt1/3, i.e. time intervals from Dstmin up to (1/2)Dstmin
%(Dt1/2 = t((1/2)Dstmin) _ t(Dstmin)) and (1/3)Dstmin (Dt1/3 =
%t((1/3)Dstmin) _ t(Dstmin)), respectively. Comparison of
%two data sets corresponding to Dt1/2 and Dt1/3 allows us to
%make conclusions about dynamics of Dst index during storm
%recovery phase. 
Figure 1 schematically shows method of calculation of durations of main and recovery phases.

\section{Results}

\subsection{Average values of durations} 

Tables 1 and 2 present number of storms, average values and standard deviations 
for durations of main ($\Delta T$) and recovery ($\Delta t_{1/2}$ and 
$\Delta t_{1/3}$) phases of magnetic storms induced by various types of solar wind streams 
(CIR, Ejecta, MC, Sheath and IND) as well as the ratios of these durations %(DT/Dt1/2 and DT/Dt1/3) 
$\Delta T/\Delta t_{1/2}$ and $\Delta T/\Delta t_{1/3}$ 
In order to show a possible influence of value of storms ($Dst_{min}$) on durations 
the results are presented separately for all,  moderate and strong magnetic storms.   

Despite a wide spread (standard deviation) of data it is possible to note a tendency 
that for compression regions (Sheath and CIR) the duration of main phase is less, 
and duration of recovery phase is more than for ICME (magnetic clouds and Ejecta), 
thus the ratios of durations for compression regions are less than 0.5 for $\Delta t_{1/2}$ 
(0.3 for $\Delta t_{1/3}$), and for ICME more than 0.5 (0.3). 
This tendency is observed both for the strong, and for moderate storms. 

\subsection{Relations between main phase durations and $Dst_{min}$ } 

Figure 2 shows dependences of main phase durations $\Delta T$ on storm value $Dst_{min}$ 
for magnetic storms 
induced by various types of solar wind streams when data are selected in 25 nT bins of $Dst_{min}$. 
In order o compare our results with data published by Hutchinson et al (2011), 
we calculate parameters for a values 'ICME" which is sum of Sheath, MC and Ejecta and 
is similar to "CME" used in paper by Hutchinson et al (2011). 
Taking large spread of data into account we can suggest that there is a slight increase of duration 
$\Delta T$ in region of (-125 … -100 nT) only for MC. 
There is no increasing for other types of interplanetary drivers including 'ICME" (
in contrast with results by Hutchinson et al (2011). 

Figure 3 shows dependences of storm value $Dst_{min}$ on main phase durations $\Delta T$ 
for magnetic storms induced by various types of solar wind streams 
when data are selected in 3-h bins of $\Delta T$. 
This figure demonstrate that at short durations of main phase all drivers induce moderate storms 
with close value of $Dst_{min}$ while at longer durations Sheath and MC generate stronger storms 
than CIR and Ejecta. 

\subsection{Relations between durations of two phases }

In contrast with data of Tables 1 and 2 Figures 4-6 present individual durations of 2 phases 
for Sheath, CIR and complex MC+Ejecta, respectively. Left panels present data for 
$\Delta t_{1/2}$, right panels for $\Delta t_{1/3}$, 
while upper, second  and bottom panels for strong, moderate and all storms, respectively. 
Number of storms is presented in right panels. 

Figures 4 and 5 shows that there is anticorrelation between magnetic storms induced by CIR and Sheath. 
This anticorrelation is clearer for stronger storms and for shorter $\Delta t_{1/2}$ duration, 
and it is absent for moderate and all storms with $\Delta t_{1/3}$. 
Dependence between magnetic storms induced by body of CME (MC+Ejecta) is absent or slight reverse one

\section{Discussion} 

Obtained results show that there are 2 classes of interplanetary drivers induced similar dynamics of magnetic storms: 
compressed regions (CIR and Sheath) and body of MC and Ejecta. 
In contrast with body MC and Ejecta the compressed regions are very disturbed regions and 
have higher density (dynamic pressure), temperature, Alfvenic Mach number, and 
variations of these plasma parameters and IMF. 
Thus, the similar reaction of magnetosphere on similar driver is natural 
but reason of its different reaction on types of 2 different classes is unknown now, 
and this problem require further investigations.  

One of the possible mechanisms of generation of magnetic storms is formation of new ion and 
electron radiation belts during magnetosphere compression by interplanetary shocks (see, e.g., 
\citep{Lorentzenetal2002,Tverskayaetal2003,ObaraLi2003,Hudsonetal2004,Looperetal2005,BorovskyDenton2006},
and references therein). 
Both Sheath and CIR can have shock in front of them and compress magnetosphere. 
So, faster fall of $Dst$ index during main phase and slower change of $Dst$ index 
during recovery phase of storms induced by them may be connected 
with formation of new radiation belts. 
%Less good fittings for strong storms generated by CIR and Sheath than by MC and Ejects (Figure 2) may be explained by the
%same cause. 

\section{Conclusions} 

Thus we separated various large-scale types of solar wind streams and found interplanetary drivers
for 572 magnetic storms on the basis of OMNI data of plasma and IMF parameters of solar wind 
during 1976-–2000. These data allowed us to compare temporal evolution of Dst index during main and 
recovery phases of magnetic storms induced by CIR, Sheath and body of ICME
(including MC and Ejecta). 
Our study allowed to obtain the following results.

1. Durations of both main and recovery phases depend on types of interplanetary drivers.  

2. Average durations of main phase of magnetic storms induced by CIR and Sheath are shorter 
than ones induced by MC and Ejecta. 

3. Ratios of average durations of main phase and recovery phase at level of $1/2 Dst_{min}$ 
(respectively $1/3 Dst_{min}$) are less 0.5 (0.3) for Sheath- and CIR-induced storms 
while they are larger 0.5 (0.3) for MC- and Ejecta-induced storms.

4. Anticorrelation between  durations of main and recovery phases is observed for magnetic storms 
induced by CIR and Sheath. 
This anticorrelation is stronger for stronger storms and for duration of recovery phase 
at level of $1/2 Dst_{min}$ than one at $1/3 Dst_{min}$.

\begin{acknowledgments}
The authors are grateful for the opportunity to use the OMNI database. The OMNI data were obtained from 
GSFC/ SPDF OMNIWeb (http://omniweb.gsfc.nasa.gov). This work was supported by the 
Russian Foundation for Basic Research, projects 10–02–00277a and 13-02-00158, and by 
Program 22 of Presidium of the Russian Academy of Sciences.
\end{acknowledgments}

\end{article}

%% Enter Figures and Tables here:

% When submitting articles through the GEMS system:
% COMMENT OUT ANY COMMANDS THAT INCLUDE GRAPHICS.
%
% FOR FIGURES, DO NOT USE \psfrag or \subfigure commands.
%
% Figure captions go below the figure.
% Table titles go above tables; all other caption information
%  should be placed in footnotes below the table.

% DRAFT figure/table, including eps graphics
%
% \begin{figure}
% \noindent\includegraphics[width=20pc]{samplefigure.eps}
% \caption{Caption text here}
% \end{figure}
% \end{document}
%
% \begin{table}
% \caption{}
% \end{table}
%
% ---------------
% TWO-COLUMN figure/table
%
% \begin{figure*}
% \noindent\includegraphics[width=39pc]{samplefigure.eps}
% \caption{Caption text here}
% \end{figure*}
%
% \begin{table*}
% \caption{Caption text here}
% \end{table*}

% Figure 1 

\begin{figure}
\noindent\includegraphics[width=10cm]{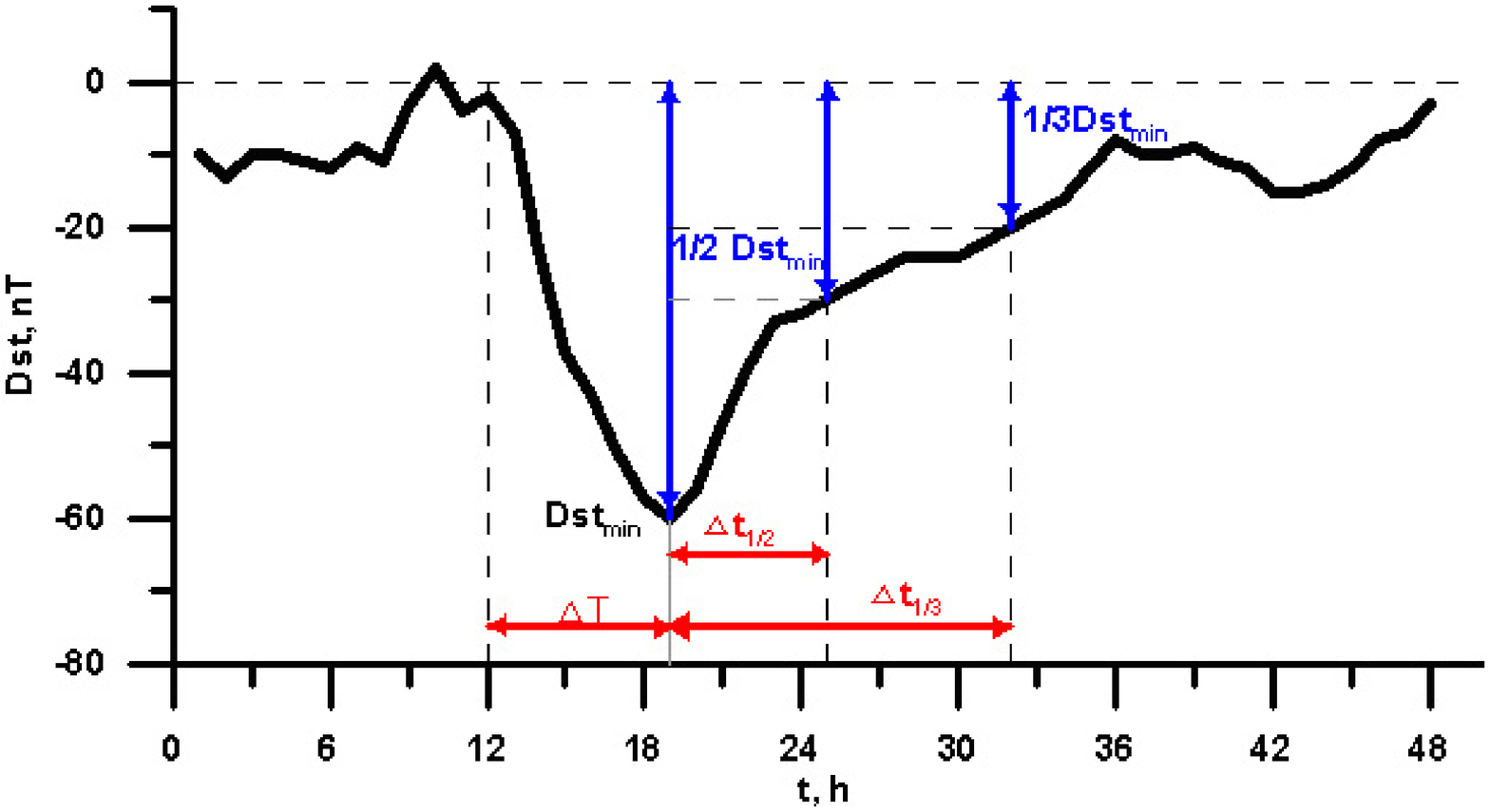}
\caption{Schematic view of method for calculation of durations of main and recovery phases for magnetic storms}
\end{figure}

% Figure 2

\begin{figure}
\noindent\includegraphics[width=10cm]{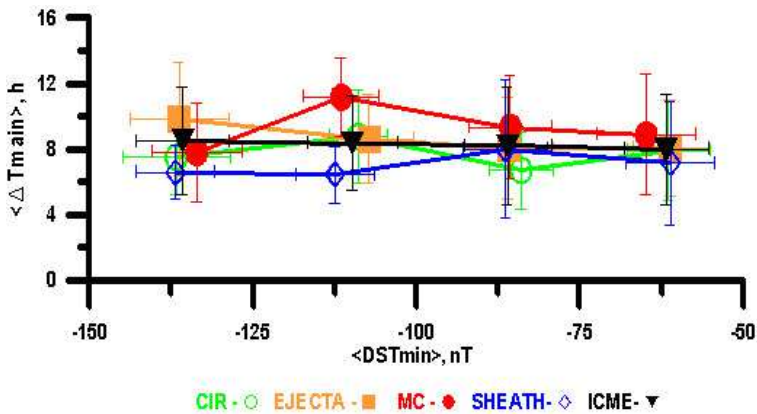}
\caption{Dependence of the main phase duration on the value of storm $Dst_{min}$  for different types of solar wind.}
\end{figure}

% Figure 3

\begin{figure}
\noindent\includegraphics[width=10cm]{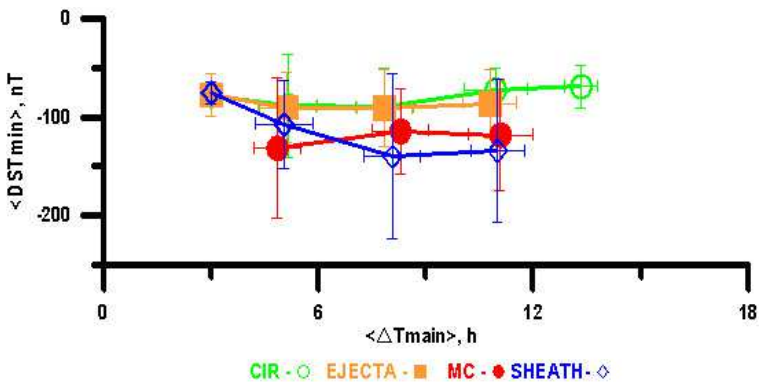}
\caption{Dependence of storm value $Dst_{min}$ on the main phase duration for different types of solar wind.}
\end{figure}

% Figure 4

\begin{figure}
\noindent\includegraphics[width=10cm]{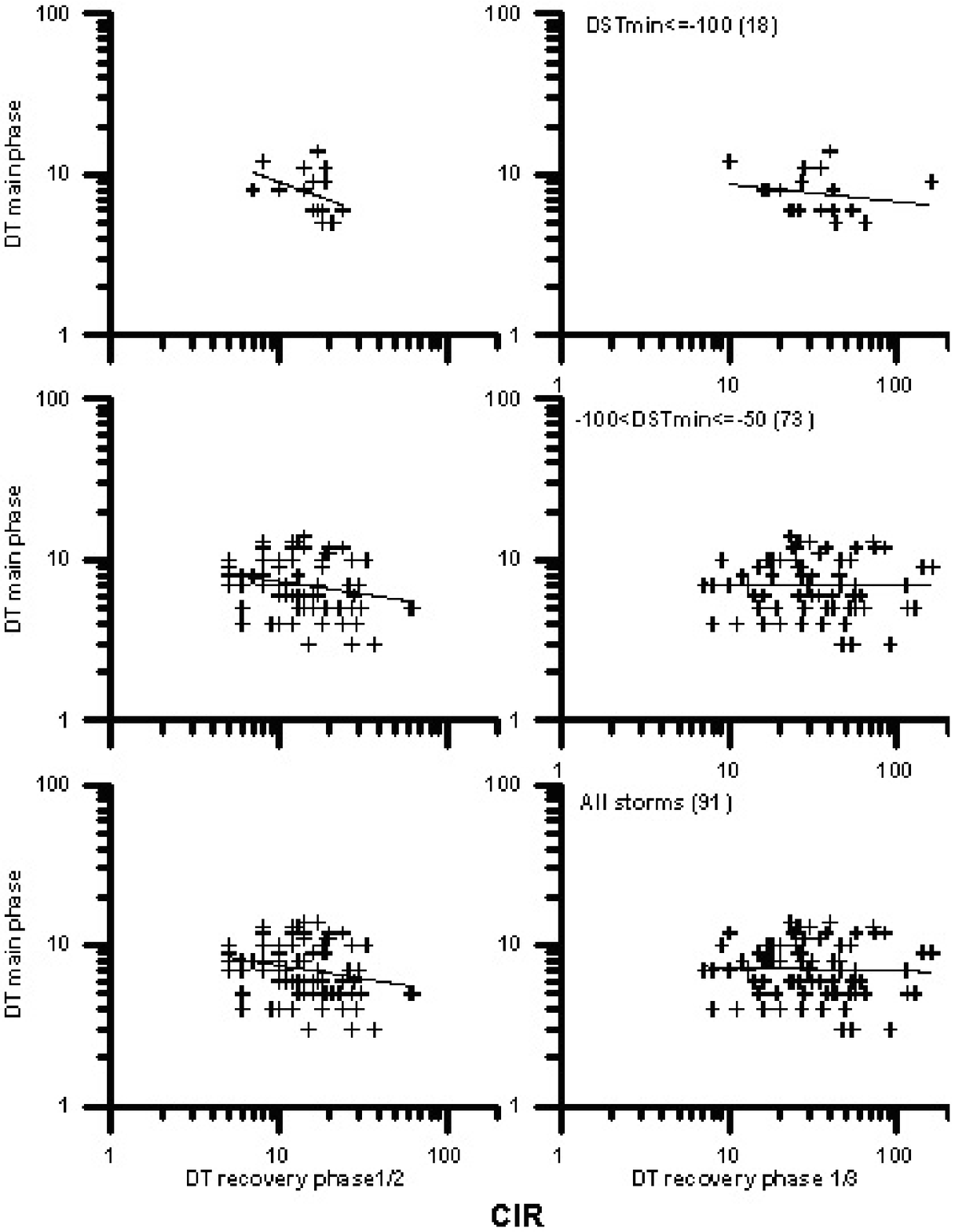}
\caption{Dependence of the main phase duration on duration of recovery phase for magnetic storms driven by CIR.}
\end{figure}

% Figure 5
\begin{figure}
\noindent\includegraphics[width=12cm]{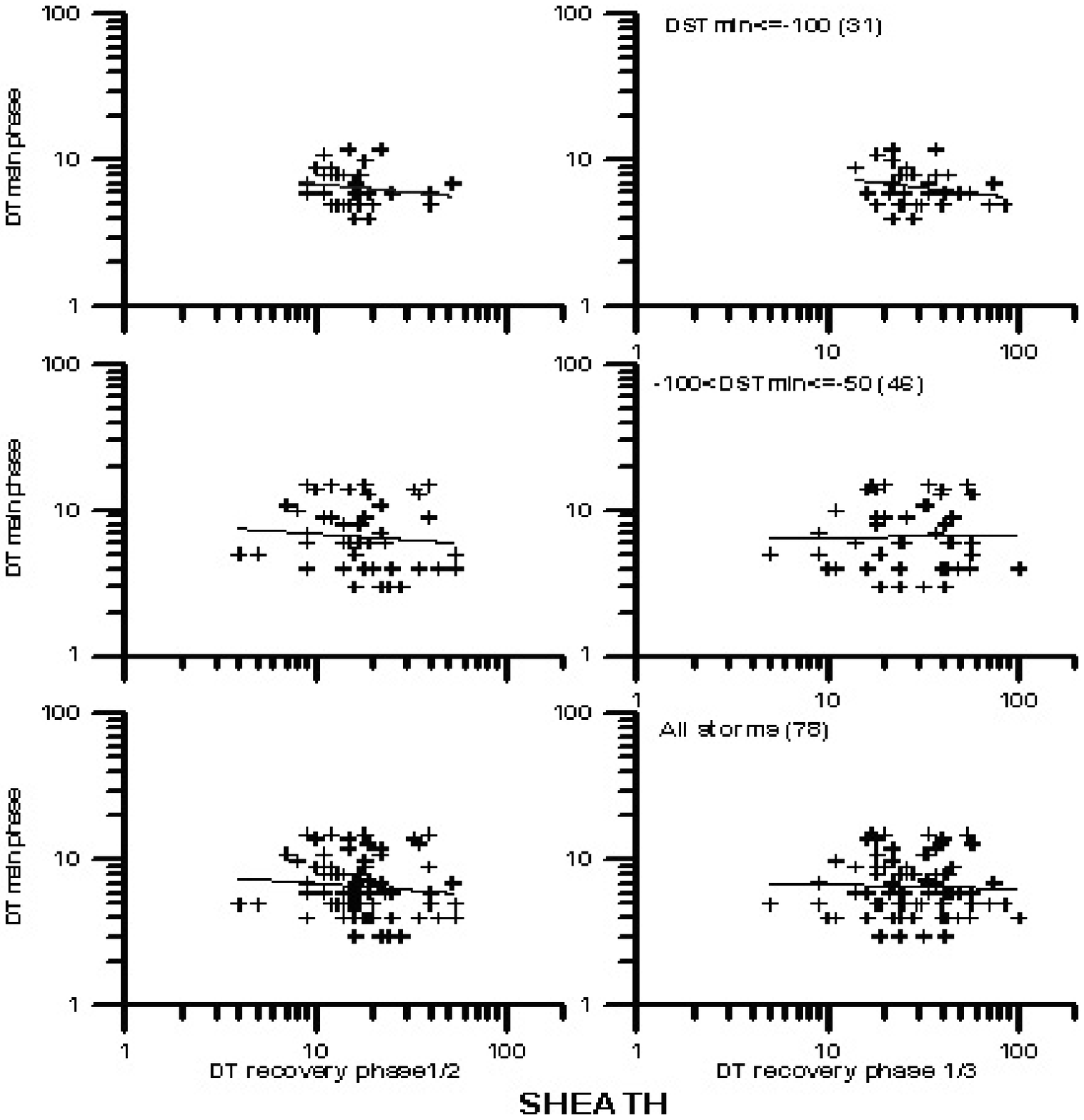}
\caption{Dependence of the main phase duration on duration of recovery phase for magnetic storms driven by Sheath 
}
\end{figure}

% Figure 6
\begin{figure}
\noindent\includegraphics[width=12cm]{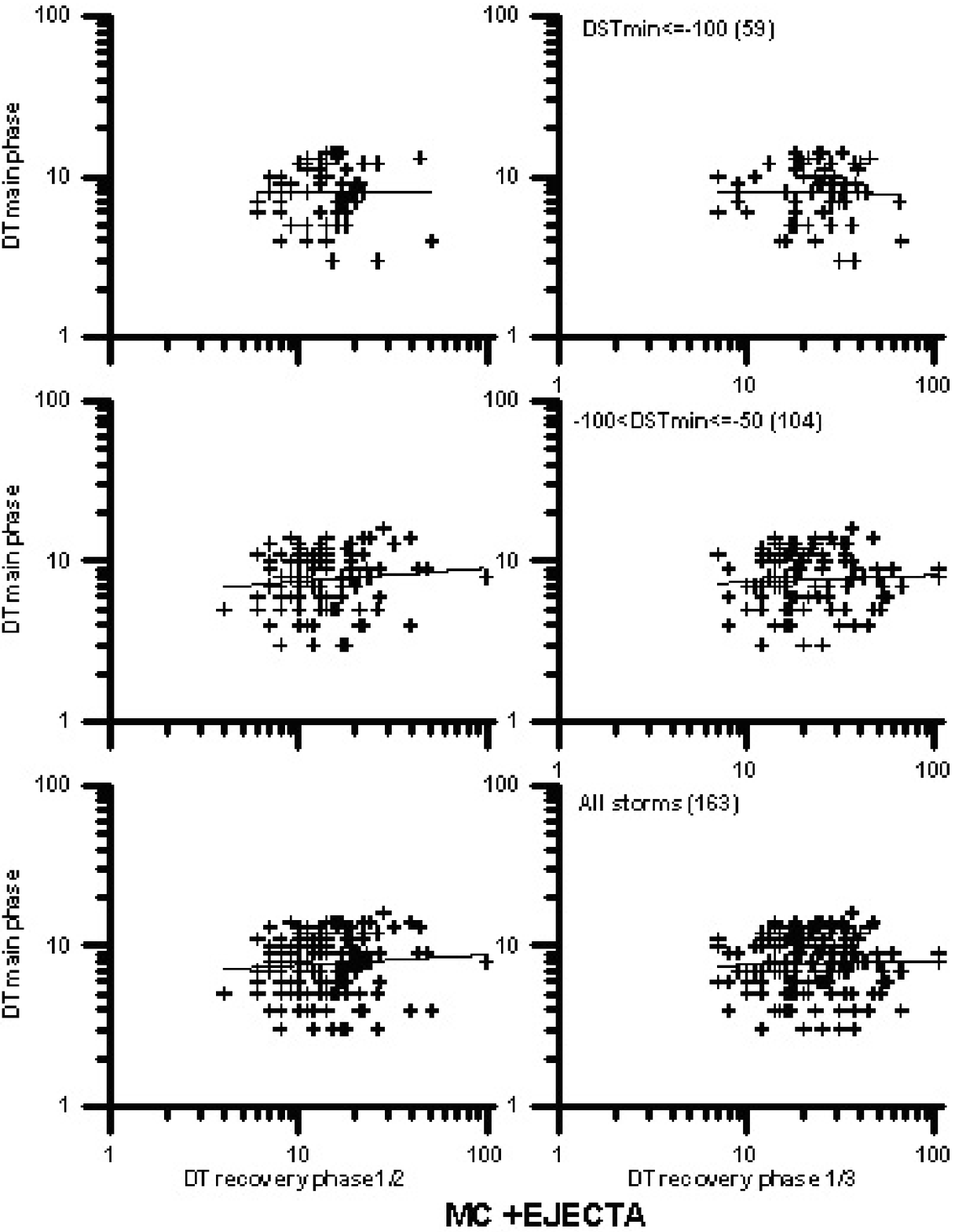}
\caption{Dependence of the main phase duration on duration of recovery phase for magnetic storms driven by SMC+Ejecta}
\end{figure}

% ---------------
% EXAMPLE TABLE
%
%\begin{table}
%\caption{Average values and standard deviations of recovery phase durations.}
%\centering
%\begin{tabular}{l c}
%\hline
% Run  & Time (min)  \\
%\hline
%  $l1$  & 260   \\
%  $l2$  & 300   \\
%  $l3$  & 340   \\
%  $h1$  & 270   \\
%  $h2$  & 250   \\
%  $h3$  & 380   \\
%  $r1$  & 370   \\
%  $r2$  & 390   \\
%\hline
%\end{tabular}
%\tablenotetext{a}{Footnote text here.}
%\end{table}

\begin{table}
{\small 
\caption{Average duration of main phase, duration of recovery phase at level of 1/2 $Dst_{min}$ and their ratio for various types of solar wind streams and value of storm.
}
\centering
\begin{tabular}{l|cccc|cccc|cccc}
\hline
SW   &  \multicolumn{4}{c}{$All storms$} & 
               \multicolumn{4}{|c}{Stoms with $-100<Dst \le -50$}  &
	\multicolumn{4}{|c}{Storms with $Dst \le -100$}  \\
%	\multicolumn{2}{|c}{$Dst \le -150$}  &
%	\multicolumn{2}{|c}{$Dst \le -200$}  \\
%             & \multicolumn{2}{c}{$\Delta t_{1/3}$\tablenotemark{a}} & 
%               \multicolumn{2}{|c}{$\Delta t_{1/2}$\tablenotemark{b}}  \\
\cline{2-13}
%& &&&&&&&&&& \\
%type      &  Number & $<\Delta T>_{main}$ & $<\Delta T>_{rec}$ & ratio & Number & $<\Delta T>_{main}$ & $<\DeltaT>_{rec}$ & ratio & Number & $<\DeltaT>_{main}$ & $<\DeltaT>_{rec}$ & ratio \\  
type      &  N & $<\Delta T>$ & $<\Delta t_{1/2}>$ & ratio & N & $<\Delta T>$ & $<\Delta t_{1/2}>$ & ratio 
& N & $<\Delta T>$ & $<\Delta t_{1/2}>$ & ratio \\
\hline
  CIR                 & 91      & $7.7 \pm 2.9$  & $16.1 \pm 9.9$    & 0.48       & 73   & $7.6 \pm 2.9$  &  $16.1 \pm 10.9$  & 0.47    & 18 & $8.0 \pm 2.5$ & $15.9 \pm 4.2$ & 0.5    \\
  Ejecta            & 114    & $8.3 \pm 3.3$  & $15.4 \pm 10.5$  & 0.54       & 80   & $8.1 \pm 3.0$  &  $15.1 \pm 11.7$  & 0.54    & 34 & $8.7 \pm 3.1$ & $16.0 \pm 3.1$ & 0.54  \\
  MC                 & 49      & $8.7 \pm 3.4$  & $16.0 \pm 9.4$    & 0.54       & 24   & $9.0 \pm 3.5$  &  $17.8 \pm 10.1$  & 0.5      & 25 & $8.4 \pm 3.2$ & $14.2 \pm 8.2$ & 0.59  \\
  Sheath           & 77      & $7.3 \pm 3.4$  & $19.0 \pm 3.4$    & 0.38       & 46   & $7.6 \pm 4.0$  &  $19.8 \pm 11.8$  & 0.38    & 31 & $6.9 \pm 2.2$ & $17.7 \pm 9.5$ & 0.39  \\
  IND                 & 266   & $8.4 \pm 3.3$  & $16.8 \pm 9.5$    & 0.5         & 202  & $8.2 \pm 3.3$  &  $17.3 \pm 10.3$  & 0.47   & 64 & $9.0 \pm 3.1$ & $15.3 \pm 6.2$ & 0.59  \\
\hline
\end{tabular}
%\tablenotetext{a}{$\Delta t_{1/2} = t(1/2 Dst_{min}) - t(Dst_{min})$}
%\tablenotetext{b}{$\Delta t_{1/3} = t(1/3 Dst_{min}) - t(Dst_{min})$}
}
\end{table}

\begin{table}
{\small
\caption{Same as in Table 1 for duration of recovery phase at level of 1/3 $Dst_{min}$.
}
\centering
\begin{tabular}{l|cccc|cccc|cccc}
\hline
SW   &  \multicolumn{4}{c}{$All storms$} & 
               \multicolumn{4}{|c}{Stoms with $-100<Dst \le -50$}  &
	\multicolumn{4}{|c}{Storms with $Dst \le -100$}  \\
%	\multicolumn{2}{|c}{$Dst \le -150$}  &
%	\multicolumn{2}{|c}{$Dst \le -200$}  \\
%             & \multicolumn{2}{c}{$\Delta t_{1/3}$\tablenotemark{a}} & 
%               \multicolumn{2}{|c}{$\Delta t_{1/2}$\tablenotemark{b}}  \\
\cline{2-13}
%& &&&&&&&&&& \\
%  type      &  Number & $<\Delta T>_{main}$ & $<\Delta T>_{rec}$ & ratio & Number & $<\Delta T>_{main}$ & $<\DeltaT>_{rec}$ & ratio & Number & $<\DeltaT>_{main}$ & $<\DeltaT>_{rec}$ & ratio \\
type      &  N & $<\Delta T>$ & $<\Delta t_{1/3}>$ & ratio & N & $<\Delta T>$ & $<\Delta t_{1/3}>$ & ratio 
& N & $<\Delta T>$ & $<\Delta t_{1/3}>$ & ratio \\
\hline
  CIR                 & 91      & $7.7 \pm 2.9$  & $39.7 \pm 31.6$  & 0.19        & 73    & $7.6 \pm 2.9$  &  $39.8 \pm 31.5$  & 0.19     & 18 & $8.0 \pm 2.5$ & $39.3 \pm 32.3$ & 0.2  \\
  Ejecta            & 114    & $8.3 \pm 3.3$  & $26.6 \pm 16.5$  & 0.31        & 80    & $8.1 \pm 3.0$  &  $26.2 \pm 18.2$  & 0.31     & 34 & $8.7 \pm 3.1$ & $27.6 \pm 11.5$ & 0.32  \\
  MC                 & 49      & $8.7 \pm 3.4$  & $24.6 \pm 12.2$  & 0.35        & 24    & $9.0 \pm 3.5$  &  $25.4 \pm 13.4$  & 0.35     & 25 & $8.4 \pm 3.2$ & $23.7 \pm 11.4$ & 0.35  \\
  Sheath           & 77      & $7.3 \pm 3.4$  & $32.8 \pm 17.8$  & 0.22        & 46    & $7.6 \pm 4.0$  &  $32.3 \pm 18.3$  & 0.24     & 31 & $6.9 \pm 2.2$ & $33.6 \pm 17.0$ & 0.21  \\
  IND                 & 266   & $8.4 \pm 3.3$  & $33.3 \pm 22.7$  & 0.25         & 202 & $8.2 \pm 3.3$  &  $34.5 \pm 23.4$  & 0.24     & 64 & $9.0 \pm 3.1$ & $29.6 \pm 20.0$ & 0.3  \\
\hline
\end{tabular}
%\tablenotetext{a}{$\Delta t_{1/2} = t(1/2 Dst_{min}) - t(Dst_{min})$}
%\tablenotetext{b}{$\Delta t_{1/3} = t(1/3 Dst_{min}) - t(Dst_{min})$}
}
\end{table}

\end{document}